\documentclass{elsarticle}
\usepackage{amsmath}
\usepackage{amsfonts}
\usepackage[utf8]{inputenc}
\usepackage{hyperref}
\usepackage{graphicx}
\usepackage{tikz}
\newcommand{\ket}[1]{| #1 \rangle}
\newcommand{\bra}[1]{\langle #1 |}
\DeclareUnicodeCharacter{03BB}{$\lambda$}
\DeclareUnicodeCharacter{2009}{\,}
\newcommand{\urlalt}[2]{\href{#2}{\nolinkurl{#1}}}

\makeatletter
\def\ps@pprintTitle{%
 \let\@oddhead\@empty
 \let\@evenhead\@empty
 \def\@oddfoot{}%
 \let\@evenfoot\@oddfoot}
\makeatother

\begin{document}
\title{A tutorial introduction to quantum circuit programming in dependently typed Proto-Quipper\tnoteref{t1}}
% \titlerunning{A tutorial introduction to dependently typed
% Proto-Quipper}
% \author{Peng Fu\inst{1} \and
% Kohei Kishida\inst{2} \and
% Neil J. Ross\inst{1} \and Peter Selinger\inst{1}}
\tnotetext[t1]
{This work was supported by the Air Force Office of Scientific Research
under award number FA9550-15-1-0331. Any opinions, findings and
conclusions or recommendations expressed in this material are those of
the authors and do not necessarily reflect the views of the
U.S. Department of Defense.
}
\author[1]{Peng Fu}
\author[2]{Kohei Kishida}
\author[1]{Neil J. Ross}
\author[1]{Peter Selinger}

\address[1]{Dalhousie University, Halifax, NS, Canada}
\address[2]{University of Illinois, Urbana-Champaign, IL, U.S.A.}
% \address[3]{Dalhousie University, Halifax, NS, Canada}
% \address[4]{Dalhousie University, Halifax, NS, Canada}

% \address{Dalhousie University, Halifax, NS, Canada
%   \email{\{frank-fu,neil.jr.ross,peter.selinger\}@dal.ca} \and
%   University of Illinois, Urbana-Champaign, IL, U.S.A.\\
%   \email{kkishida@illinois.edu}
% }

\begin{abstract}
We introduce dependently typed Proto-Quipper, or
Proto-Quipper-D for short, an experimental quantum circuit programming
language with linear dependent types. We give several examples to
illustrate how linear dependent types can help in the construction of
correct quantum circuits. Specifically, we show how dependent types
enable programming families of circuits, and how dependent types solve
the problem of type-safe uncomputation of garbage qubits. We also
discuss other language features along the way.

\end{abstract}
\begin{keyword}
  Quantum programming languages \sep Linear dependent types \sep Proto-Quipper-D
\end{keyword}
\maketitle
\section{Introduction}

Quantum computers can in principle outperform conventional computers
at certain crucial tasks that underlie modern computing
infrastructures. Experimental quantum computing is
 in its early stages and existing devices are not yet suitable for practical
computing. However, several groups of researchers, in both academia and industry, are
now building quantum computers (see, e.g., \cite{google,ibm,umd}).
Quantum computing also raises many challenging questions for the
programming language community \cite{mosca_et_al}: How should we
design programming languages for quantum computation? How should we
compile and optimize quantum programs? How should we test and verify
quantum programs? How should we understand the semantics of quantum
programming languages?

In this paper, we focus on quantum circuit programming using the
linear dependently typed functional language Proto-Quipper-D.

The no-cloning property of quantum mechanics states that one cannot in
general copy the state of a qubit. Many existing quantum programming
languages, such as
Quipper\cite{green2013introduction,green2013quipper}, QISKit
\cite{qiskit}, Q\# \cite{qsharp}, Cirq \cite{cirq}, or ProjectQ
\cite{projectq}, do not enforce this property. As a result,
programmers have to ensure that references to qubits within a program
are not duplicated or discarded.  Linear types have been used for
resource aware programming \cite{girard1987linear,wadler1990linear}
and it is now well-known that they can be used to enforce no-cloning
\cite{selinger2005lambda}.  A variety of programming languages use
linear types for quantum circuit programming, e.g., Proto-Quipper-S
\cite{ross2015algebraic}, Proto-Quipper-M \cite{RiosS17}, and QWire
\cite{paykin2017qwire}.  All well-typed programs in these
languages satisfy the no-cloning property.

Dependent types \cite{martin1984intuitionistic} have been one of the main focuses in programming language and type system research in the past
decades. Dependent types make it possible to express program invariants and constraints using types
\cite{agda,bove2008dependent,coq}. In
the context of quantum circuit programming, dependent types are useful for expressing
parameterized families of circuits. For example, one can define a function that inputs a size and outputs a circuit of the corresponding size. Because the type of the output circuit is indexed by the size argument, errors due to an attempt to compose mismatched circuits are detected at compile time. Another important application of dependent types is the type-safe management of garbage qubits, which we discuss in Section \ref{exists}.

We introduce an experimental quantum circuit programming language
called dependently typed Proto-Quipper, or Proto-Quipper-D for
short. Following Quipper, Proto-Quipper-D is a
functional language with quantum data types and aims to
provide high-level abstractions for constructing quantum
circuits. Like its predecessors Proto-Quipper-S and Proto-Quipper-M,
the Proto-Quipper-D language relies on linear types to enforce
no-cloning. 
Proto-Quipper-D additionally features the use of linear dependent types
to facilitate the type-safe construction of circuit
families \cite{FKS2020}. This paper provides a practical introduction to programming
in Proto-Quipper-D.

The paper is structured around several programming examples
that showcase the use of linear dependent types in Proto-Quipper-D.
\begin{itemize}
\item We give an introduction to
  dependent types by showing how to use them to
  prove basic properties of addition in Section~\ref{dep}.

\item We show how to program with families of quantum circuits in Section~\ref{circuits}.

\item We give a new application of existential dependent types and
  show how it simplifies the construction of certain reversible quantum circuits in Section~\ref{exists}.

\end{itemize}
An implementation of Proto-Quipper-D is available at: \url{https://gitlab.com/frank-peng-fu/dpq-remake}.
All the programs in
this tutorial are available at \texttt{test/Tutorial.dpq} in the Proto-Quipper-D repository.

\section{An introduction to dependent types}
\label{dep}
Proto-Quipper-D supports programming by recursion and pattern matching. For example, the following is
a program that defines the addition of Peano numbers.

\begin{verbatim}
data Nat = Z | S Nat

add : !(Nat -> Nat -> Nat)
add n m =
  case n of
    Z -> m
    S n' -> S (add n' m)
\end{verbatim}

In the above program, we use the keyword \texttt{data} to define an
algebraic data type in the style of Haskell 98 \cite{jones2003haskell}.
The type checker will analyze the data type declaration and determine
that \texttt{Nat} is a \textit{parameter type} (or \textit{non-linear type}).
In Proto-Quipper-D,
parameter types are types that can be freely duplicated and discarded.
The addition function has type \texttt{!(Nat -> Nat -> Nat)}. The exclamation mark
(pronounced ``bang'') in front
of a function type makes that type a parameter type. This means that addition is a reusable function, i.e., it can be used multiple times. The type of a non-reusable function
would be of the form \texttt{a -> b} and in particular would not be prefixed by a ``\texttt{!}''. In contrast to a reusable function, a non-reusable function must be used exactly once. This guarantees that any quantum data embedded in the function does not get inadvertently duplicated or discarded. Proto-Quipper-D
requires all top-level declarations to have parameter types, making them reusable.

With dependent types, we can even encode properties of programs in
types. In Proto-Quipper-D, dependent function types are of the form
\texttt{(x : A) -> B}, where the type \texttt{B} may optionally
mention the variable \texttt{x}. We can think of this dependent
function type as the universal quantification $\forall x:A\,.\,B$ of
predicate logic. Dependent types therefore allow us to represent
properties of programs as types. For example, the following programs
correspond to proofs of basic properties of addition.

\begin{verbatim}
addS : ! (p : Nat -> Type) -> (n m : Nat) -> 
             p (add n (S m)) -> p (add (S n) m) 
addS p n m h = 
  case n of
    Z -> h
    S n' -> addS (λ y -> p (S y)) n' m h

addZ : ! (p : Nat -> Type) -> (n : Nat) -> p (add n Z) -> p n
addZ p n h = case n of
               Z -> h
               S n' -> addZ (λ y -> p (S y)) n' h
\end{verbatim}

\noindent The type of \texttt{addS} expresses the theorem that for all
natural numbers $n$ and $m$, we have $n + S m = S n + m$. However,
rather than using an equality symbol, we use the so-called
\textit{Leibniz equality}. Leibniz defined two things to be equal if
they have exactly the same properties. Therefore, the type of
\texttt{addS} states that for any property \texttt{p : Nat -> Type}
of natural numbers, and for all natural numbers \texttt{n},
\texttt{m}, if \texttt{add n (S m)} has the property \texttt{p}, then
\texttt{add (S n) m} has the property \texttt{p}.  Similarly, the type
of \texttt{addZ} expresses the fact that $n + Z = n$.

Note how the types of dependent type theory play a dual role: on the
one hand, they can be read as types specifying the inputs and outputs
of functional programs; on the other hand, they can be read as logical
statements. This is the so-called \textit{propositions-as-types}
paradigm {\cite{GLT89}}. For example, the last arrow ``\texttt{->}''
in the type of \texttt{addS} can be interpreted both as a function
type and as the logical implication symbol. This works because a proof
of an implication is actually a function that transforms evidence for
the hypothesis into evidence for the conclusion.

Indeed, not only does the type of the function \texttt{addS}
corresponds to a theorem, but the actual code of \texttt{addS}
corresponds to its proof. For example, in the branch when \texttt{n}
is \texttt{Z}, the variable \texttt{h} has type \texttt{p (add Z (S
  m))}, which equals \texttt{p (S m)} by the definition of
\texttt{add}.  This branch is expecting an expression of type
\texttt{p (add (S Z) m)}, which equals \texttt{p (S m)} by definition
of \texttt{add}, so the type-checking of \texttt{h} succeeds.

In practice, we often use the above equality proofs to convert
one type to another. We will give examples of this in
Section~\ref{conv}. However, we emphasize that Proto-Quipper-D is designed for
quantum circuit programming, not general theorem proving like languages such as Coq and Agda.
The only kind of primitive propositions we can have are equalities, 
and the support of dependent data types 
is limited to \textit{simple types}, as discussed in Section \ref{simple}.

\section{Programming quantum circuits}
\label{circuits}
We use the keyword \texttt{object} to introduce simple linear objects
such as bits and qubits, representing primitive wires in circuits. We
use the keyword \texttt{gate} to introduce a primitive gate. As far as
Proto-Quipper-D is concerned, gates are uninterpreted; they simply
represent basic boxes that can be combined into circuits. Each
primitive gate has a type specifying its inputs and outputs.

\begin{verbatim}
object Qubit
object Bit

gate H : Qubit -> Qubit  
gate CNot : Qubit -> Qubit -> Qubit * Qubit
gate C_X : Qubit -> Bit -> Qubit * Bit
gate C_Z : Qubit -> Bit -> Qubit * Bit
gate Meas : Qubit -> Bit
gate Discard : Bit -> Unit
gate Init0 : Unit -> Qubit
gate Term0 : Qubit -> Unit
\end{verbatim}

\noindent The above code declares primitive types \texttt{Qubit} and
\texttt{Bit} and a number of gates. For example, the gate \texttt{H}
is a reusable linear function of type \texttt{!(Qubit -> Qubit)},
which, by convention, represents the Hadamard gate. Note that the type
checker automatically adds the \texttt{!} to gate declarations, so it
is not necessary to do so manually. The type expression \texttt{Qubit
  * Qubit} denotes the tensor product of two qubits, and thus, the
controlled-not gate \texttt{CNot} has two inputs and two outputs
(where, by convention, the first input is the target and the second is
the control). By linearity, the arguments of the \texttt{CNot} can
only be used once.  Thus, an expression such as \texttt{CNot x x} will
be rejected by the type checker because the argument \texttt{x} is
used twice. We also introduce two
classically-controlled gates \texttt{C\_X} and \texttt{C\_Z}.
The gate \texttt{Meas} corresponds to a measurement,
turning a qubit into a classical bit.
The type \texttt{Unit}
represents the unit of the tensor product, i.e., a bundle of zero
wires. Thus, the gate \texttt{Discard} can be used to discard a
classical bit.

\paragraph{Qubit initialization and termination}
We use the gate \texttt{Init0} for the initialization of a qubit in
the state $\ket{0}$, and the gate \texttt{Term0} for the termination
of a qubit in the state $\ket{0}$. The gate \texttt{Term0} is the
adjoint of the gate \texttt{Init0} and we think of these gates as
being each other's inverse. If one thinks of a quantum circuit with
$n$ input qubits and $m$ output qubits as a $2^{m} \times 2^{n}$
complex matrix, then qubit initialization is modelled as the matrix
\[
\ket{0} =  \begin{pmatrix}
    1 \\
    0
  \end{pmatrix},
\]
while qubit termination is modelled as the matrix
\[
  \begin{pmatrix}
    1 & 0 \\
  \end{pmatrix} = \bra{0}. 
\]
We use the following diagrams for the gates \texttt{Init0} (on the
left) \texttt{Term0} (on the right), as they are suggestive of the
duality between these operations.
\[
  \includegraphics[scale=1.5]{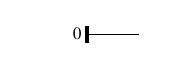}
  \includegraphics[scale=1.5]{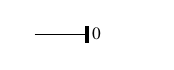}
\]

The following program produces a circuit that generates a Bell state:
\[
  \includegraphics[scale=1.5]{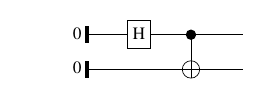}
\]

\begin{verbatim}
bell00 : !(Unit -> Qubit * Qubit)
bell00 u = 
  let x = Init0 ()
      y = Init0 ()
      x' = H x
      (y, x') = CNot y x'   
  in (y, x')
\end{verbatim}

\noindent If we want to display the
circuit generated by the function \texttt{bell00}, we can use
Proto-Quipper's \texttt{box} function:

\begin{verbatim}
bell00Box : Circ(Unit, Qubit * Qubit)
bell00Box = box Unit bell00
\end{verbatim}

\noindent The \texttt{box} function inputs a circuit-generating
function such as \texttt{bell00} and produces a completed
circuit of type \texttt{Circ(Unit, Qubit * Qubit)}. In the
Proto-Quipper-D interactive shell, we can then type \texttt{:d
  bell00Box} to display the circuit.

The following program implements quantum teleportation.
\[
  \includegraphics[scale=1.5]{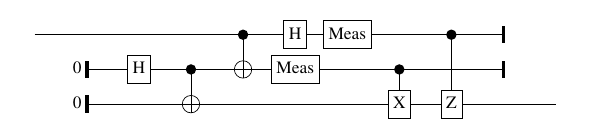}
\]
\begin{verbatim}
bellMeas : !(Qubit -> Qubit -> Bit * Bit)
bellMeas x y = 
  let (x', y') = CNot x y
      y'' = H y'
  in (Meas x', Meas y'') 

tele : !(Qubit -> Qubit)
tele phi =
  let (bob, alice) = bell00 ()
      (a', phi') = bellMeas alice phi
      (bob', a'') = C_X bob a'
      (r, phi'') = C_Z bob' phi'
      u = Discard phi''
      u = Discard a''
  in r         
\end{verbatim}

\subsection{Simple types}
\label{simple}
Following Quipper, Proto-Quipper-D makes a distinction between
\textit{parameters} and \textit{states}. Parameters are values that
are known at circuit generation time, while states are only known at
circuit execution time. For example, the type \texttt{Nat} represents
a parameter, while the type \texttt{Qubit} represents a state.

In Proto-Quipper-D, we use the concept of \textit{simple types} to
describe states. As discussed earlier, simple types can be introduced using the keyword
\texttt{object}. In practice, it is more common to
create simple types by composing existing ones.  For example, \texttt{Qubit *
  Qubit} is also a simple type. For this reason, we call the tensor
product a \textit{simple type constructor}.  In Proto-Quipper-D, the
programmer can also define families of new simple types using the
\texttt{simple} keyword.  For example, the following defines a type
family \texttt{Vec}, and \texttt{Vec Qubit n} is a simple type.

\begin{verbatim}
simple Vec a : Nat -> Type where
    Vec a Z = VNil
    Vec a (S n) = VCons a (Vec a n)
\end{verbatim}

\noindent The expression \texttt{Nat -> Type} is a \textit{kind
  expression}. It means that \texttt{Vec a n} is a type whenever
\texttt{n} is a natural number. The two clauses after the
\texttt{simple} keyword are the definition of the type \texttt{Vec a
  n}. The first clause says that an element of the type \texttt{Vec a
  Z} can be constructed by the constructor \texttt{VNil}. The second
clause says that an element of the type \texttt{Vec a (S n)} can be
constructed by applying the constructor \texttt{VCons} to a term of
type \texttt{a} and a term of type \texttt{Vec a n}. Therefore,
\texttt{Vec a n} represents a vector of \texttt{n} elements
of type \texttt{a}.

The type \texttt{Vec a n} is an example of \textit{dependent data
  type}, where the data type \texttt{Vec a n} depends on some term
\texttt{n} of type \texttt{Nat}. In the interpreter, we can query the
types of \texttt{VNil} and \texttt{VCons} (for example, by typing \texttt{:t VNil}). They have the following
types.

\begin{verbatim}
VNil : forall (a : Type) -> Vec a Z  
VCons : forall (a : Type) -> forall (n : Nat) -> 
              a -> Vec a n -> Vec a (S n)
\end{verbatim}

\noindent In Proto-Quipper-D, all data constructors are reusable, so
there is no need for them to have an explicit bang-type. The leading
\texttt{forall} keyword means that programmers do not need to supply
that argument when calling the function.  We call such quantification
\textit{irrelevant quantification}.  For example, when using
\texttt{VCons}, we only need to give it two arguments, one of type
\texttt{a} and one of type \texttt{Vec a
  n}. 

The simple data type declaration is currently the only way to introduce dependent data types
in Proto-Quipper-D. Semantically, simple types corresponds to states. Syntactically,
a simple type can uniquely determine the size and the constructors of its data.  
The type checker will check whether a simple data type declaration is well-defined. 
Note that not all dependent data types are simple types. For example, the
following declaration will not pass the type checker.

\begin{verbatim}
simple ColorVec a : Nat -> Type where
  ColorVec a Z = CNil
  ColorVec a (S n) = VConsBlue a (ColorVec a n)
  ColorVec a (S n) = VConsRed a (ColorVec a n)
\end{verbatim}
\noindent The \texttt{ColorVec} data type is ambiguous when the parameter is \texttt{S n}, as the
constructor in this case can be either \texttt{VConsBlue} or \texttt{VConsRed}.
The following is another ill-formed example.

\begin{verbatim}
simple InfVec a : Nat -> Type where
  InfVec a Z = INil
  InfVec a (S n) = IVCons a (InfVec a (S n))
\end{verbatim}

\noindent The \texttt{InfVec} data type is not a simple data type because
from the type \texttt{InfVec a (S n)}, we can not determine the size of its
data, for example, the value of the type \texttt{InfVec a (S Z)} does not have
a finite size.

In general, checking whether a simple type is well-defined is equivalent
to deciding whether a general recursive function is well-defined
and terminating, which is undecidable.
Currently, Proto-Quipper-D checks whether a simple data
type declaration is well-defined using the
same criterion as checking primitive recursion~\cite{kleene1968introduction}.

\subsection{Using Leibniz equality}
\label{conv}

Suppose we want to define a function that reverses the order of the components in a vector. One way to do this is to use an
accumulator: we traverse the vector while prepending each element to the
accumulator. This can be expressed by the \texttt{reverse\_aux} function defined below. 

\begin{verbatim}
reverse_aux : ! (a : Type) -> (n m : Nat) ->
                      Vec a n -> Vec a m -> Vec a (add n m)
reverse_aux a n m v1 v2 =
  case n of
    Z -> let VNil = v1 in v2
    S n' ->
      let VCons q qs = v1 in
      let ih = reverse_aux a n' (S m) qs (VCons q v2) in
      addS (Vec a) n' m ih
\end{verbatim}
Note that the type of \texttt{reverse\_aux} indicates that the length of the output vector
is the sum of the lengths of the input vectors. In the definition for \texttt{reverse\_aux}, we use \texttt{v1}
and \texttt{v2} exactly once in each branch, which respects linearity. In the second
branch of \texttt{reverse\_aux},
the type checker expects an expression of type \texttt{Vec a (add (S n') m)}, but
the expression \texttt{ih}, obtained from the recursive call, has type
\texttt{Vec a (add n' (S m))}. We therefore use the theorem \texttt{addS} from Section~\ref{dep} to convert the type to \texttt{Vec a (add (S n') m)}.
We can then use \texttt{reverse\_aux} to define the \texttt{reverse\_vec} function, which requires a similar type conversion. 

\begin{verbatim}
reverse_vec : ! (a : Type) -> (n : Nat) -> Vec a n -> Vec a n
reverse_vec a n v = addZ (Vec a) n (reverse_aux a n Z v VNil)
\end{verbatim}

\subsection{Families of quantum circuits}

We can use simple data types such as vectors to define functions that correspond to families of circuits. 
As an example, we consider the well-known quantum Fourier transform \cite{nielsen2002quantum}. The quantum Fourier transform is the map defined by
{\small
\[
\ket{a_1, \dots , a_n} \mapsto \frac{(\ket{0} + e^{2\pi i 0.a_1 a_2 ... a_n}\ket{1}) \dots (\ket{0} + e^{2\pi i 0.a_{n-1} a_n} \ket{1}) (\ket{0} + e^{2\pi i 0.a_n}\ket{1})}{2^{n/2}}.
\]
}where $0.a_1...a_n$ is the binary fraction $a_1/2+a_2/4 + ... + a_n /2^n$. Circuits for the quantum Fourier transform can be constructed using the Hadamard gate $H$ and the controlled rotation gates $R(k)$ defined by
\[
R(k)=\begin{pmatrix}
  1 & 0 & 0 & 0 \\
  0 & 1 & 0 & 0 \\
  0 & 0 & 1 & 0 \\
  0 & 0 & 0 & e^{2\pi i /2^k}
\end{pmatrix}.
\]
The family of gates $R(k)$ can be declared in Proto-Quipper-D as follows:
\begin{verbatim}
gate R Nat : Qubit -> Qubit -> Qubit * Qubit
\end{verbatim}
Applying the Hadamard gate to the first qubit produces the following state
\[
H_1 \ket{a_1,\dots, a_n} =  \frac{1}{\sqrt{2}}(\ket{0} + e^{2\pi i 0.a_1}\ket{1})\otimes \ket{a_2,\dots, a_n},
\]
where the subscript on the gate indicates the qubit on which the gate
acts. We then apply a sequence of controlled rotations using the the first qubit as the target. This yields
{\small
\[R(n)_{1, n} \dots R(2)_{1, 2}  H_1 \ket{a_1,\dots, a_n} =
  \frac{1}{2^{1/2}}(\ket{0} + e^{2\pi i 0.a_1 a_2 ... a_n}\ket{1}) \otimes \ket{a_2,\dots, a_n}, 
\]}where the subscripts $i$ and $j$ in $R(k)_{i, j}$ indicate the target and control qubit, respectively. When $n = 5$, the above sequence of gates corresponds to the following circuit.

\begin{center}
  \includegraphics[scale=1]{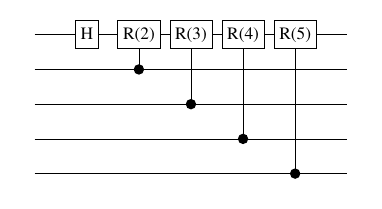}
\end{center}
To construct such a circuit in Proto-Quipper-D, we first define the \texttt{rotate} function, which will produce a cascade of rotations with a single target. The rotations in the above circuit are then generated by \texttt{oneRotation 4}.

\begin{verbatim}
rotate : ! forall (y : Nat) -> Nat -> 
                Qubit -> Vec Qubit y -> Qubit * Vec Qubit y
rotate k q v =
    case v of
      VNil -> (q, VNil)
      VCons x xs ->
        let (q', x') = R k q x
            (q'', xs') = rotate (S k) q' xs
        in (q'', VCons x' xs')

oneRotation : ! (n : Nat) -> 
                   Circ(Qubit * Vec Qubit n, Qubit * Vec Qubit n)
oneRotation n =
  box (Qubit * Vec Qubit n) 
    (λ x -> let (q, v) = x in rotate 2 (H q) v)
\end{verbatim}
The \texttt{rotate} function uses the input vector \texttt{v} for controls and recursively applies the rotation gate \texttt{R} to the target qubit \texttt{q}, updating the rotation angle at each step. To program the full quantum Fourier transform, we apply the Hadamard and controlled rotations recursively to the rest of input qubits.
\begin{verbatim}
qft : ! forall (n : Nat) -> Vec Qubit n -> Vec Qubit n
qft v =
  case v of
    VNil -> VNil
    VCons q qs ->
      let q' = H q
          (q'', qs') = rotate 2 q' qs
          qs'' = qft qs'
      in VCons q'' qs''

qftBox : ! (n : Nat) -> Circ(Vec Qubit n, Vec Qubit n)
qftBox n = box (Vec Qubit n) qft
\end{verbatim}
For example, \texttt{qftBox 5} generates the following circuit.
\begin{center}
  \includegraphics[scale=1]{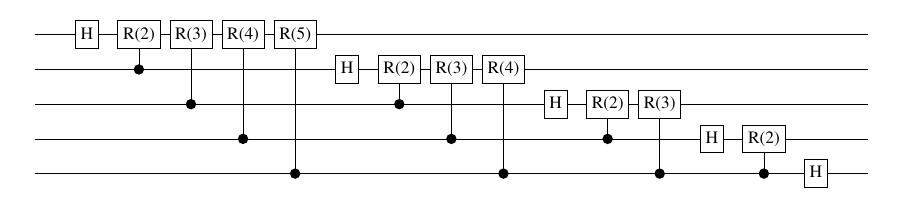}
\end{center}
The input qubits of the circuit above use a big-endian ordering. We can convert to little-endian ordering by reversing the input vector.
{\small
\begin{verbatim}
qftBoxLittle : ! (n : Nat) -> Circ(Vec Qubit n, Vec Qubit n)
qftBoxLittle n = box (Vec Qubit n) (λ v -> qft (reverse_vec Qubit n v))
\end{verbatim}
}
\noindent Then \texttt{qftBoxLittle 5} generates the following circuit.
\begin{center}
  \includegraphics[scale=1]{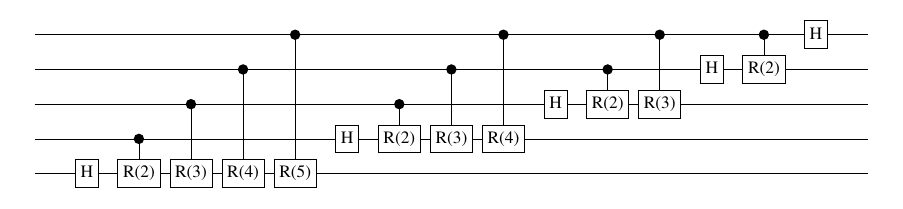}
\end{center}

\subsection{Type classes for simple types and parameter types}

Proto-Quipper-D is equipped with a type class mechanism that allows the user to define type classes and instances~\cite{wadler1989make}. In addition, Proto-Quipper-D has two built-in type classes called
\texttt{Simple} and \texttt{Parameter}, which are useful for programming with simple types and parameter
types, respectively. The user cannot directly define instances for these two classes. Instead,
instances for \texttt{Simple} and \texttt{Parameter} are automatically generated
from data type declarations.

When a simple data type is defined, the type checker automatically makes the type an instance of the \texttt{Simple} class and, if appropriate, of the \texttt{Parameter} class. Similarly, when algebraic data types such as \texttt{List} and \texttt{Nat} are defined, the type checker makes instances of the \texttt{Parameter} class when possible. For example, consider the following programs.

\begin{verbatim}
data List a = Nil | Cons a (List a)

kill : ! forall a -> (Parameter a) => a -> Unit
kill x = ()

test1 : !(List Nat -> Unit)
test1 x = kill x

test2 : !(List Qubit -> Unit)
test2 x = kill x
\end{verbatim}
The argument of the function \texttt{kill} must be a parameter. The
expression \texttt{test1} is well-typed, because \texttt{List Nat} is
a member of the \texttt{Parameter} class. But \texttt{test2} fails to
type-check because \texttt{List Qubit} is not a member of the
\texttt{Parameter} class.

Simple types are useful for describing the types of certain operations
that require a circuit, rather than a family of circuits. Examples are
boxing, unboxing, and reversing a circuit:

\begin{verbatim}
box : (a : Type) -> forall (b : Type) ->
          (Simple a, Simple b) => !(a -> b) -> Circ(a, b)  

unbox : forall (a b : Type) ->
           (Simple a, Simple b) => Circ(a, b) -> !(a -> b)

reverse : forall (a b : Type) ->
            (Simple a, Simple b) => Circ(a, b) -> Circ(b, a)
\end{verbatim}
The type of \texttt{box} implies that
only functions of simple type can be turned into boxed 
circuits. The following program will not type-check because \texttt{List Qubit} is not a simple type.

\begin{verbatim}
boxId : Circ(List Qubit, List Qubit)
boxId = box (List Qubit) (λ x -> x)
\end{verbatim}
With the built-in function \texttt{reverse}, we can now compute the inverse of
\texttt{qftBox}.
\begin{verbatim}
boxQftRev : ! (n : Nat) -> Circ(Vec Qubit n, Vec Qubit n)
boxQftRev n = reverse (qftBox n)
\end{verbatim}
By definition, the family of circuits represented by
\texttt{boxQftRev} is obtained by taking the inverse of every member
of the family of circuits represented \texttt{qftBox}. For example,
\texttt{boxQftRev 5} generates the following circuit.
 
\begin{center}
  \includegraphics[scale=1]{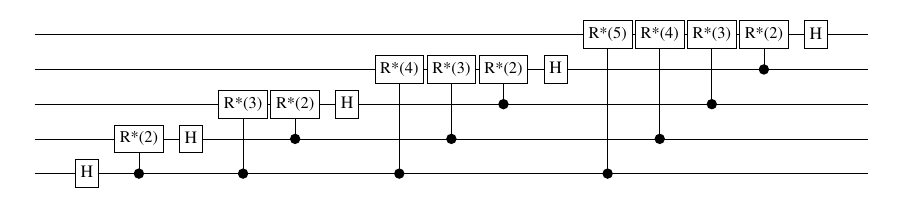}
\end{center}

\section{Type-safe management of garbage qubits}
\label{exists}

In quantum computing, it is often necessary to provide classical oracles to a quantum algorithm. These oracles are reversible implementations of classical boolean functions. Consider the example of the single bit full adder. If the inputs are \texttt{a}, \texttt{b} and \texttt{carryIn}, then the boolean expression \texttt{xor (xor a b) carryIn} calculates the sum
of \texttt{a, b} and \texttt{carryIn} while the boolean expression \texttt{(a \&\& b) || (a \&\& carryIn) || (b \&\& carryIn)} calculates the output carry.

We can implement the single bit adder as a reversible quantum circuit.
Suppose that the boolean operations \texttt{xor}, \texttt{||}, and
\texttt{\&\&} are given as reversible circuits of type \texttt{!(Qubit
  -> Qubit -> Qubit * Qubit)}. Here, the first qubit in the output of
each function is the result of the operation, whereas the second qubit
is a ``garbage'' qubit that cannot be discarded since this would
violate linearity. As a result, the following naive implementation of
the adder generates 7 garbage qubits and has a 9-tuple of qubits as
its return type.

\begin{verbatim}
adder : ! (Qubit -> Qubit -> Qubit -> 
              Qubit * Qubit * Qubit * Qubit * Qubit *
                  Qubit * Qubit * Qubit * Qubit)  
adder a b carryIn = 
  let (a1, a2, a3) = copy3 a
      (b1, b2, b3) = copy3 b
      (carryIn1, carryIn2, carryIn3) = copy3 carryIn
      (g1, r) = xor a1 b1
      (g2, s) = xor carryIn1 r
      (g3, c1) = a2 && b2
      (g4, c2) = a3 && carryIn2
      (g5, c3) = b3 && carryIn3
      (g6, c4) = c1 || c2
      (g7, carryOut) = c4 || c3
  in (s, carryOut, g1, g2, g3, g4, g5, g6, g7)
\end{verbatim}

\noindent Due to linearity, the copying of a classical qubit must be explicit. In the code above, \texttt{copy3} is a function that produces
three copies of a qubit that is in a classical state, i.e., \texttt{copy3} corresponds to the following circuit.

\begin{center}
  \includegraphics[scale=1]{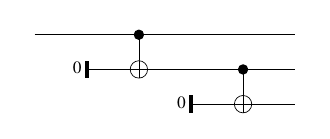}
\end{center}

The above implementation of the adder is hard to read and awkward to compose with other circuits, because its type keeps track of all the garbage qubits produced throughout the computation. In Proto-Quipper-D, we solve this problem using monads \cite{jones1995functional}, existential
dependent types, and existential circuit boxing.

Instead of using the type \texttt{!(Qubit -> Qubit -> Qubit * Qubit)}, we give \texttt{xor}, \texttt{||}, and \texttt{\&\&} the type \texttt{!(Qubit -> Qubit -> WithGarbage Qubit)}, where \texttt{WithGarbage} is a monad that will take care of the garbage qubits. The idiomatic implementation of the full adder in Proto-Quipper-D is the following.
{\small
  %% Note: the following verbatim environment contains some Unicode
  %% characters U+2009 ("thin space").
\begin{verbatim}
adder : !(Qubit * Qubit * Qubit -> WithGarbage (Qubit * Qubit))
adder input = do
  let (a, b, carryIn) = input
      (a1, a2, a3) = copy3 a
      (b1, b2, b3) = copy3 b
      (carryIn1, carryIn2, carryIn3) = copy3 carryIn 
  s <- [| xor (xor a1 b1) (pure carryIn1)|]
  carryOut <- [| [| (a2 && b2) || (a3 && carryIn2) |] || (b3 && carryIn3) |]
  return (s, carryOut)
\end{verbatim}
}
\noindent
Proto-Quipper-D implements
idiom brackets \cite{mcbride2008applicative} of the form \texttt{[| f a b c |]}. This expression
will be translated to \texttt{join (ap (ap (ap (pure f) a) b) c)}, where
\texttt{ap}, \texttt{pure} and \texttt{join} have the following types. 

\begin{verbatim}
ap : ! forall (a b : Type) -> forall (m : Type -> Type) ->
           (Monad m) => m (a -> b) -> m a -> m b

pure : ! forall (m : Type -> Type) ->
           (Monad m) => forall (a : Type) -> a -> m a

join : ! forall (a : Type) -> forall (m : Type -> Type) ->
             (Monad m) => m (m a) -> m a
\end{verbatim}

\noindent We now briefly discuss the definition of the \texttt{WithGarbage} monad.

\begin{verbatim}
data WithGarbage a = WG ((n : Nat) * Vec Qubit n) a

instance Monad WithGarbage where
  return x = WG (Z, VNil) x
  bind wg f = let WG ng r = wg
                  (n, g) = ng
                  WG mg' r' = f r
                  (m, g') = mg'
               in WG (add n m, append g g') r'
\end{verbatim}

\noindent The type \texttt{(x : A) * B} is an \emph{existential
  dependent type}, corresponding to the existential quantification
$\exists x:A\,.\,B$ of predicate logic. Just as for dependent function
types, the type $B$ may optionally mention the variable $x$. The elements of the type
\texttt{(n : Nat) * Vec Qubit n} are pairs \texttt{(n, v)}, where
\texttt{n : Nat} and \texttt{v : Vec Qubit n}. Thus,
\texttt{WithGarbage a} contains a vector of qubits of a unknown length
and a value of type \texttt{a}. In the definition of the
\texttt{WithGarbage} monad, the \texttt{return} function does not
generate any garbage qubits. The \texttt{bind} function combines the
garbage qubits from the two computations \texttt{wg} and \texttt{f}.
Note that it uses the \texttt{append} function to concatenate two vectors.

The standard way to dispose of a qubit (and turn it into garbage) is via
the following \texttt{dispose} method.
\begin{verbatim}
class Disposable a where
  dispose : a -> WithGarbage Unit

instance Disposable Qubit where
  dispose q = WG (1, VCons q VNil) ()
\end{verbatim}

\noindent So for example, we can implement \texttt{xor} as follows. Note
that the implemented circuit is not optimal, but it serves to
illustrate the point.

\begin{verbatim}
xor : !(Qubit -> Qubit -> WithGarbage Qubit)
xor x y =
  do let z = Init0 ()
         (z', x') = CNot z x
         (z'', y') = CNot z' y
     dispose x'
     dispose y'
     return z''
\end{verbatim}

Using the \texttt{WithGarbage} monad, we can essentially program as if the
extra garbage qubits do not exist. Next, we need a type-safe way to
uncompute the garbage qubits. We achieve this with the function
\texttt{with\_computed} below, which takes a garbage-producing
function and turns it into a function that produces no garbage.  The
implementation of \texttt{with\_computed} relies on the following
built-in function:
\begin{verbatim}
existsBox : (a : Type) -> forall (b : Type) -> 
               (Simple a, Parameter b) => (p : b -> Type) -> 
               !(a -> (n : b) * p n) ->
               (n : b) * ((Simple (p n)) => Circ(a, p n))
\end{verbatim}
Intuitively, the \texttt{existsBox} construct is used to box an
existential function. It takes a circuit generating function of type
\texttt{!(a -> (n : b) * p n)} as input and turns it into an
\textit{existential circuit} of the type \texttt{(n : b) * Circ(a, p
  n)}. Using \texttt{existsBox}, we can define
\texttt{with\_computed}:

{\small
  %% Note: the following verbatim environment contains some Unicode
  %% characters U+2009 ("thin space").
\begin{verbatim}
with_computed : !forall d -> {a b c : Type} -> (Simple a, Simple b) =>
                  !(a -> WithGarbage b) -> !(c * b -> d * b) ->
                   (c * a -> d * a)
with_computed a b c f g input =
  let (y, x) = input
      (_, circ) =
         existsBox a (\x -> Vec Qubit x * b)
           (\z -> unGarbage (f z))
      h' = unbox circ
      (v, r) = h' x
      circ_rev = unbox (reverse circ)
      (d, r') = g (y, r)
      res = circ_rev (v, r')
  in (d, res)
\end{verbatim}
}

\noindent 
The \texttt{with\_computed} function inputs a function
\texttt{f : !(a -> WithGarbage b)} and a second function \texttt{g : !(c *
  b -> d * b)}, and produces a garbage-free circuit \texttt{c * a -> d
  * a} corresponding to the following diagram. Of course each wire may
correspond to multiple qubits, as specified in its type. The notation \texttt{\{a b c : Type\} -> ...}
means that when using the \texttt{with\_computed} function, we do not need to
supply the type arguments \texttt{a, b, c}, they will be inferred by the compiler. 

\[
  \begin{tikzpicture}[scale=0.8]
    \draw (-3.5,2) -- node[above] {$c$} (-0.5,2);
    \draw (0.5,2) -- node[above] {$d$} (3.5,2);
    \draw (-1.5,1) -- node[above] {$b$} (-0.5,1);
    \draw (0.5,1) -- node[above] {$b$} (1.5,1);
    \draw (-3.5,1) -- node[above] {$a$} (-2.5,1);
    \draw (2.5,1) -- node[above] {$a$} (3.5,1);
    \draw (-2,0) -- node[above] {garbage} (2,0);
    \draw[fill=white] (-2.5,-0.3) rectangle node {$f$} (-1.5,1.3);
    \draw[fill=white] (1.5,-0.3) rectangle node {$f^{-1}$} (2.5,1.3);
    \draw[fill=white] (-0.5,0.7) rectangle node {$g$} (0.5,2.3);
  \end{tikzpicture}
\]

\noindent Note that this construction is type-safe, because it
guarantees that there will be no uncollected garbage, regardless of
how much garbage the function $f$ actually produces. However,
Proto-Quipper-D does not guarantee the \emph{semantic} correctness of
the resulting circuit; it could happen that a qubit that is supposed
to be returned in state $\ket{0}$ is returned in some other
state. Since semantic correctness is in general undecidable,
Proto-Quipper-D makes no attempt to prove it. Consequently, a failure
of semantic correctness is considered to be a programming error,
rather than a type error. However, the \emph{syntactic} correctness of
the generated circuits is guaranteed by the type system.

Using the \texttt{with\_computed} function and a few helper functions,
we can obtain the following \texttt{adderRev} circuit, where all the
garbage qubits are uncomputed.

\begin{center}
  \includegraphics[scale=0.35]{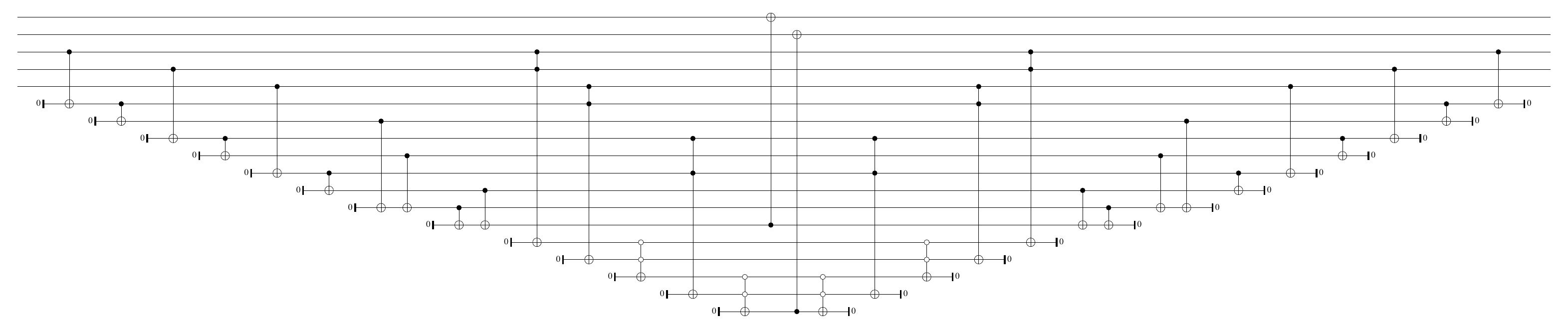}
\end{center}

\begin{verbatim}
adderRev : Circ(Qubit * Qubit * (Qubit * Qubit * Qubit),
                Qubit * Qubit * (Qubit * Qubit * Qubit))
adderRev =
   box (Qubit * Qubit * (Qubit * Qubit * Qubit))
      (\ x -> with_computed adder copy x)
\end{verbatim}
\section{Case studies}

Beyond the simple examples that were considered in this tutorial, we
have conducted two nontrivial programming case studies using
Proto-Quipper-D. The first one is an implementation of the binary
welded tree algorithm {\cite{Childs-et-al-2003}}, which features the use of
the dependent vector data type. The second is a boolean oracle for
determining the winner of a completed game of Hex, which
features the use the of \texttt{WithGarbage} and \texttt{State}
monads. Both implementations are distributed with Proto-Quipper-D, in
\texttt{test/BWT.dpq} and \texttt{test/Hex3.dpq}, respectively. The
largest oracle contains 457,383 gates. For this oracle, type checking
is nearly instantaneous (it takes less than 1 second), and circuit
generation takes about 2.5 minutes on a 3.5 GHz CPU (4 cores), 16 GB
memory desktop machine.

\section{Related work}

Proto-Quipper encompasses a family of languages that aim to provide a formal foundation
for the Quipper quantum programming language. Both Quipper and Proto-Quipper are 
functional programming languages that make the distinction between parameters (values known
at circuit generation time) and states (values known at circuit execution time). The main difference
between Proto-Quipper and Quipper is that Proto-Quipper features linear and dependent
types to help ensure the correctness of the constructed circuits.

Many other quantum programming languages exist. Here, we briefly discuss
a few of them. 
IBM's Qiskit \cite{qiskit} is a framework for quantum computing based on Python.
Qiskit comes with simulator backends and API for IBM's quantum computers. The basic workflow of
Qiskit is somewhat similar to that of Quipper: the programmer imports the necessary Qiskit libraries in Python and then composes quantum circuits (by constructing a circuit object).
The programmer can then test and analyze circuits by running them against the appropriate backend.   

Microsoft's Q\# \cite{qsharpdoc}
is a quantum programming language that features classically controlled
programs that interact with quantum states. Q\# makes a distinction between \textit{operations} and
\textit{functions}, where operations instruct a quantum machine (or simulator) to perform computations,
while functions are just classical procedures that are executed by a classical computer. Q\# has
built-in primitive types such as array, qubit, boolean. Furthermore, Q\# supports tuples and struct-like
user-defined types.  

Google's Cirq \cite{cirq}
is a Python library for writing, manipulating, and optimizing quantum
circuits, and for running them against quantum computers and
simulators. The Cirq language exposes more hardware details to the
programmer than Qiskit. For example, the programmer may choose the
arrangement of qubits (e.g., in a line or on a grid). Cirq supports
user-defined gates as long as certain methods in the \textit{gate
  class} are defined. Cirq allows programmers to construct quantum
circuits via a collection of \textit{moments}, i.e., a circuit time
step where many gates can be applied to different qubits at the same
time. The iteration of moments is also possible in Cirq.

\section{Conclusion}

In this tutorial, we introduced the quantum programming language
Proto-Quip\-per-D through a series of examples. Proto-Quipper-D is an
experimental language and is currently under active development.
We did not discuss all of the features of
Proto-Quipper-D, since our goal was to highlight the use of linear and
dependent types in quantum circuit programming. Finally, we also briefly discuss
some related languages from the industry.

\bibliographystyle{plain}
\bibliography{tutorial}

\end{document}